\documentclass[prl, twocolumn,superscriptaddress]{revtex4}
\usepackage{amsmath,amssymb,bm}
\usepackage{graphicx}
\usepackage{epstopdf}
\usepackage{latexsym}
\usepackage{subfigure}
\usepackage{color}
\usepackage{natbib}
\usepackage{braket}
      \usepackage{hyperref}
\hypersetup{
  colorlinks,
  citecolor=magenta,
  linkcolor=blue,
  urlcolor=blue}

\usepackage{tikz}
\newcommand{\mybox}{%
\tikz[baseline]{\draw[line width=0.4mm] (0,0) -- (0.15cm,0) -- (0.15cm,0.15cm) -- (0,0.15cm) -- (0,-0.2mm)}%
} 
 
\newcommand{\mybars}{%
\tikz[baseline]{\draw[line width=0.4mm] (-0.2mm,0) -- (0.15cm+0.2mm,0); \draw[line width=0.4mm] (-0.2mm,0.15cm) -- (0.15cm+0.2mm,0.15cm)}%
} 
 
\newcommand{\mybendl}{%
\tikz[baseline]{\draw[line width=0.4mm] (-0.2mm,0) -- (0.15cm,0) -- (0.15cm,0.15cm) -- (-0.2mm,0.15cm)}%
} 
 
\newcommand{\mybendr}{%
\tikz[baseline]{\draw[line width=0.4mm] (0.15cm+0.2mm,0.15cm) -- (0,0.15cm) -- (0,0) -- (0.15cm+0.2mm,0)}%
}

\begin{document}

\title{ Kagome model for a ${\mathbb Z}_2$ quantum spin liquid}

\author{Matthew S. Block}
\affiliation{Department of Physics \& Astronomy, California State
  University, Sacramento, CA 95819}

\author{Jonathan D'Emidio}
\affiliation{Institute of Physics, Ecole Polytechnique F\'ed\'erale de Lausanne (EPFL), CH-1015 Lausanne, Switzerland}

\author{Ribhu K. Kaul}
\affiliation{Department of Physics \& Astronomy, University of Kentucky, Lexington, KY 40506-0055}

\begin{abstract}
We present a study of a simple model antiferromagnet consisting of a
sum of nearest neighbor SO($N$) singlet projectors on the Kagome
lattice. Our model shares some features with the popular $S=1/2$
Kagome antiferromagnet but is specifically designed to be free of the
sign-problem of quantum Monte Carlo.  In our numerical analysis, we
find as a function of $N$ a quadrupolar magnetic state and a wide
range of a quantum spin liquid. A solvable large-$N$ generalization suggests that the quantum spin liquid in our
original model is a gapped ${\mathbb Z}_2$ topological phase. Supporting this assertion, a numerical
study of the entanglement entropy in the sign free model shows a quantized  topological
contribution.
\end{abstract}
\date{\today}
\maketitle

Quantum antiferromagnetism on the Kagome lattice is an
important playground in
the study of quantum spin liquids emerging from frustrated
magnetism. The most popular model in this family is the $S=1/2$ Kagome anti-ferromagnet $H = J \sum_{\braket{ij}} \vec S_i \cdot \vec S_j$. 
Despite a quarter of a century of intense research using an array of
numerical and analytic methods on this important model the ground state of
the $S=1/2$ Kagome anti-ferromagnet remains hotly contested.
While the absence of magnetic order is
uncontroversial~\cite{marston1991:kagome,chalker1992:kagome,singh1992:kagseries,leung1993:kagome,lecheminant1997:kagome}, various
nonmagnetic ground states have been proposed
including, e.g., an array of quantum spin
liquids~\cite{ran2007:kagome,yan2011:qsl,iqbal2013:kagome,he2017:kagome} and valence bond solid ordering~\cite{marston1991:kagome,nikolic2003:kagvbs,singh2007:kagvbs}. In parallel to the theoretical work, a number of synthetic
quantum materials have been identified that provide venues where the interplay of
quantum fluctuations and frustration on the Kagome lattice give rise to novel unexplained behavior~\cite{balents2010:spliq}.

Of all the proposed phases of matter on the Kagome, the so-called gapped ${\mathbb Z}_2$
quantum spin liquid~\cite{sachdev1992:kagome} is the simplest example of an exotic state with
long range entanglement~\cite{savary2016:rev}, a prototypical quantum state that
cannot be deformed into a simple product or mean field state. In its
simplest incarnation, the
excitations above the ground state come in two basic varieties, an $e$ particle
and an $m$ particle which by themselves are bosons but are mutual semions~\cite{wen1991:to,kitaev2008:lh}. Remarkably it has been shown that the presence of
these excitations can be detected in the entanglement of the ground state wavefunction
itself, giving rise to a contribution called the ``topological
entanglement entropy''~\cite{levin2006:tee,kitaev2006:tee}.  Although
this state is not yet experimentally accesible, we now have
a few model Hamiltonians that realize this topological order, including the toric
code~\cite{kitaev2003:toric}, the honeycomb Kitaev model~\cite{kitaev2008:lh}, non-bipartite quantum dimer
models~\cite{moessner2001:rvb,misguich2002:kqdm} and models of frustrated
bosons~\cite{balents2002:bfg,isakov2006:z2,dang2011:ring}. It is clearly of great interest to extend this family of models with an eye to finding
simple models that could find realizations in physical systems.

{\em Model:}  A number  of variations on the basic $S=1/2$ Heisenberg model have been
introduced and studied on the Kagome lattice, including
Sp($N$)~\cite{sachdev1992:kagome}, SU($N$)~\cite{corboz2012:kagsun},
larger spin versions of the two spin Heisenberg exchange~\cite{changlani2015:kags1}, as well as
certain multi-spin interactions~\cite{bauer2014:kagome}. 
In this work we present and study a new variant of the Kagome
anti-ferromagnet. Our model is constructed from spins which have a local Hilbert
space of $N$ states, denoted for site $j$ as $\Ket{\alpha}_j$ where
$\alpha=1,\ldots,N$. The Hamiltonian can be written simply as a sum of singlet projectors on the nearest neighbors of the Kagome lattice,
\begin{eqnarray}
\label{eqn:proj}
H&=&-J\sum_{\braket{ij}}\Ket{S_{ij}}\Bra{S_{ij}},\\
\label{eqn:sonsing}
\Ket{S_{ij}}&=&\frac{1}{\sqrt{N}}\sum_\alpha\Ket{\alpha\alpha}_{ij}.
\end{eqnarray}
 Physically, the Hamiltonian Eq.~(\ref{eqn:proj}) can be viewed as
 lowering the energy of singlet formation locally between nearest
 neighbors. Since all pairs of neighbors cannot simultaneously form
 singlets, quantum fluctuations play an important role in stabilizing
 the ground state.
 We note here that the usual $S=1/2$ Heisenberg model is also a sum of
 singlet projectors,  of the form Eq.~(\ref{eqn:proj}) but with
 $\Ket{S_{ij}} \rightarrow \frac{\Ket{\uparrow \downarrow} -
   \Ket{\downarrow \uparrow}}{\sqrt{2}}$, which aside from the crucial
 relative minus sign, is identical to our
 singlet Eq.~(\ref{eqn:sonsing}) at $N=2$. It is this discrepancy of sign that allows us to
 sidestep the infamous sign problem and carry out large volume
 numerical studies that are so far impossible for the $S=1/2$ Kagome
 Heisenberg model.

The model Eqs.~(\ref{eqn:proj},\ref{eqn:sonsing}) has a global SO($N$) symmetry in which each site transforms in the fundamental representation, $|\alpha\rangle
\rightarrow
O_{\alpha\beta}|\beta\rangle$ and in the
path integral can be interpreted as a statistical mechanics model of tightly
packed unoriented loops~\cite{kaul2012:tris1}. A previous study~\cite{kaul2015:son} on the triangular lattice found a $\sqrt{12}\times\sqrt{12}$
valence bond solid order at large values of $N$. Here by introducing a solvable large-$N$ limit and a
numerical study of the
entanglement entropy at finite-$N$, we show that the
increased geometric frustration of the Kagome lattice realizes a
${\mathbb Z}_2$ topological quantum spin liquid.

\begin{figure}[t]
\centerline{\includegraphics[width=\columnwidth]{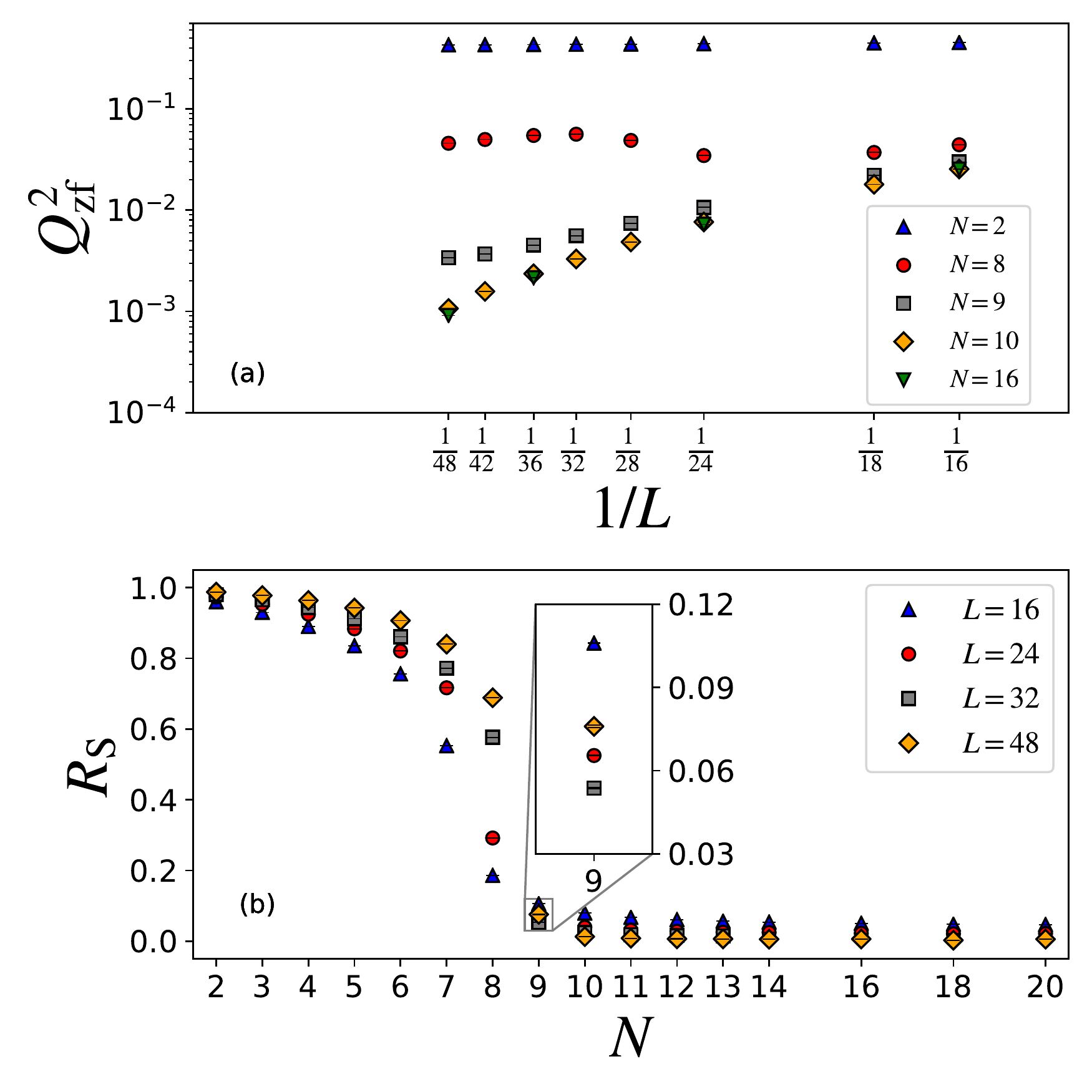}}
\caption{\label{fig:RSvsN} (color online).  Finite size scaling of the
  quadrupolar order parameter for the model Eq.~(\ref{eqn:proj}),
  shows the presence of long range order for $N \leq 9$
  and its absence for $N>9$. The upper panel shows the
  $1/L$ scaling of the order
  parameter $Q^2_{\rm zf}$. The
  lower panel shows the correlation ratio
  $R_{S}$ as a function of $N$ for different $L$.  In the thermodynamic limit a value of 1 indicates
long-range order, and 0 the absence of order. All values of $N$ show $R_S$ varying monotonically with increasing $L$ except for $N=9$.  The inset
  shows the non-monotonicity for $N=9$, where for larger systems sizes there is a
  trend of $R_S$ to increase with $L$ indicating quadrupolar
  long-range order.}
\end{figure}

We simulate the model Hamiltonian
Eqs.~(\ref{eqn:proj}-\ref{eqn:sonsing}) using the
stochastic series expansion~\cite{sandvik2010:vietri}  with loop
updates on $3\times L\times L$ lattices at an inverse temperature
$\beta$. To characterize the breaking of SO($N$) symmetry we introduce
the operator
$\hat{Q}_{\alpha\beta}=\Ket{\alpha}\Bra{\beta}-\frac{\delta_{\alpha\beta}}{N}$
which because of its tensorial nature we will call the ``quadrupolar'' order parameter. 
The Fourier transformed susceptibility,
$\chi_Q(\mathbf{k})=\frac{1}{\beta N_\text{site}}\int_0^\beta
d\tau\sum_\mathbf{r}e^{i\mathbf{k}\cdot\mathbf{r}}\langle {Q}_{\alpha\alpha}(\mathbf{r},\tau)
{Q}_{\alpha\alpha}(\mathbf{0},0)\rangle,$ is used to diagnose
quadrupolar order. We define $Q^2_{\rm zf}=\chi_Q({\bf 0})$ as 
the order parameter, a quantity which scales to a finite value in the thermodynamic limit
in the quadrupolar phase and to zero otherwise.  To facilitate detection of the long
range order, we also study a correlation ratio $R_\text{S}=1-\frac{\chi_{Q}(\mathbf{G}/L)}{\chi_{Q}(\mathbf{0})}$
where ${\bf G}$ is the shortest reciprocal lattice vector, which
scales to 1(0) in the symmetry broken (unbroken) phase. As shown in
Fig.~\ref{fig:RSvsN}, the quadrupolar order decreases as $N$ is
increased. Finite size scaling shows that for $N\leq 9$ there is
quadrupolar order that breaks the SO($N$) symmetry and for $N> 10$
the quadrupolar order vanishes. The $N=9$ case is on the verge of
transition but a careful finite size scaling indicates that it is
quadrupolar ordered. We have searched extensively for translational
symmetry breaking at the $N$ for which quadrupolar order is absent (as
was found in the triangular lattice~\cite{kaul2015:son}),
but we find no evidence for this order, indicating the possibility of
a liquid like state. We now present field theoretic arguments and
numerical evidence that this phase is a ${\mathbb Z}_2$ quantum spin liquid.

{\em Large-$N$ limit -- } To introduce a solvable large-$N$ limit that can capture both the quadrupolar as
well as non-magnetic phase, we
generalize the spins in our model to transform under larger
representations than the fundamental SO($N$), using Schwinger bosons in which
each local spin state is
associated with one of $N$ flavors of boson $b_{i\alpha}$ ( with
$[b_{i\alpha},b^\dagger_{j\beta}]=\delta_{ij}\delta_{\alpha\beta}$). The
generalized spin model is then,
\begin{eqnarray}
\label{eq:Hb}
H_b &=& -\frac{J}{N} \sum_{\langle ij \rangle} 
\left (   b^\dagger_{i\alpha}b^\dagger_{j\alpha}\right )
\left (   b^{\phantom \dagger}_{j\beta}b_{i\beta} \right )
\end{eqnarray}
with  the constraint $\sum_\alpha
b^\dagger_{i\alpha}b_{i\alpha}=n_b$, which fixes the representation of
the spin. Thus the family of models Eq.~(\ref{eq:Hb}) has two parameters
$n_b$ and $N$. Clearly $n_b=1$ corresponds to 
Eq.~(\ref{eqn:proj}). Increasing $n_b$ is a generalization of
Eq.~(\ref{eqn:proj}), with different
representations of SO($N$). These are SO($N$) analogues of the well known Schwinger boson
method of implementing higher representations of SU($N$) ~\cite{arovas1988:sb,read1989:vbs}.

\begin{figure}[t]
\centerline{\includegraphics[width=\columnwidth]{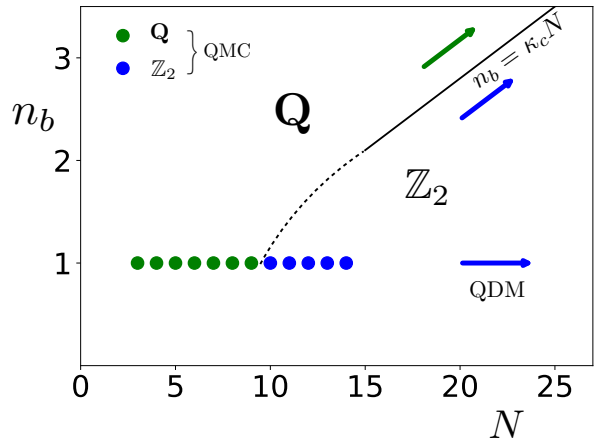}}
\caption{(color online).  The $n_b$-$N$ phase diagram obtained for 
  Eq.~(\ref{eq:Hb}) from QMC and large-$N$ limits: {\bf Q} is for
  quadrupolar and $\mathbb{Z}_2$ is the topological spin liquid. The solid circles represent 
  the QMC results from Fig.~\ref{fig:RSvsN} for $n_b=1$ which is identical to 
  Eq.~(\ref{eqn:proj}). The large-$N$ Schwinger boson gives us the 
  phase diagram when both $n_b,N\rightarrow \infty$, holding their 
  ratio $\kappa = n_b/N$ fixed. At fixed $n_b=1$ and large-$N$ a 
  quantum dimer model (QDM) on the Kagome lattice is obtained. The dashed 
  line is a guide to the eye representing the simplest way the QMC and 
large-$N$ results could be connected.}
\label{fig:nbNpd}
\end{figure}

Using boson coherent states we obtain a functional integral representation of the partition
function $Z={\rm Tr}[e^{-\beta H_b}]$,
with a field $\lambda_i (\tau)$ that enforces the on-site constraint and a
Hubbard-Stratonovich field $Q_{ij}(\tau)$ that  decouples the quartic
interaction~\cite{arovas1988:sb},
\begin{eqnarray}
Z &=& \int D\lambda DQ e^{-\int d\tau {\cal L}_b}\\
{\cal L}_b &=&  b^\dagger_{i\alpha}  \partial_\tau b^{\phantom *}_{i\alpha}
             +  \frac{N}{J}|Q_{ij}|^2 + Q^*_{ij}
             b_{i\alpha}b_{j\alpha}+ {\rm c.c.} \\
&+& \lambda_i
             (  b^\dagger_{i\alpha} b_{i\alpha}- \kappa_b N),
\end{eqnarray}
where we have set $n_b = \kappa N$. Integrating out the $b$ fields
we obtain an effective action proportional to $N$. By fixing
$\kappa$ a
large-$N$ limit can be accessed
simply by a saddle
point evaluation. Assuming space and time independent $Q$ and
$\lambda$ we find evaluating the trace over bosons: $Z=e^{-\beta V N f
}$ (where $V$ is the total number of spatial unit cells), where,
$f = \frac{1}{V}\sum_{{\bf k}\alpha}\left (\frac{z Q^2}{2J} - \lambda \left (\kappa
  + \frac{1}{2}\right ) + \frac{1}{\beta} \log 2\sinh \left
      (\frac{\beta \omega_{{\bf k}\alpha}}{2}\right ) \right)$, and
$\omega_{{\bf k}\alpha} = \sqrt{\lambda^2 - 4 Q^2 \gamma^2_{{\bf k}\alpha}}$ the
dispersion of bosons and $\gamma_{{\bf k}\alpha}$ are the three modes $\alpha=1,2,3$
of the adjacency matrix on the Kagome, $\{1,
\frac{1}{2}\left ( -1\pm \sqrt{3+2 {\rm cos}(\vec k \cdot \vec
    a_i)}\right )\}$, where $\vec a_i$ are the three shortest lattice
vectors on the triangular lattice Bravais lattice. At the saddle point
(obtained by extremizing $Q$ and $\lambda$), there are two phases one
where the $b_\alpha$ have a gap and the other where they are
condensed. The condensed phase of the $b_\alpha$ breaks the
SO($N$) symmetry and corresponds to the quadrupolar
order for the spin model, the implications of the gapped phase for the
spin model are
more subtle -- we address them below. Numerically we find the
transition between these two phases where the gap goes to zero is at $\kappa_c\approx 0.148\dots$ This gives a
phase boundary $n_b=\kappa_c N$ that we show in
Fig.~\ref{fig:nbNpd} as a solid line. Also shown in solid circles are
the phases determined from QMC for $n_b=1$. From this figure it is plausible by continuity that the quadruplar phase found in QMC corresponds to the
condensation of $b_\alpha$ ($\kappa>\kappa_c$) and the liquid like
phase in the QMC corresponds to
the state in which the $b_\alpha$ are gapped ($\kappa<\kappa_c$). We now ask what
non-magnetic state the original spin model goes into when
the $b_\alpha$ acquire a gap?  Following previous work~\cite{read1989:vbs}, the state is determined by $1/N$ fluctuations beyond mean field,
which take the structure of a U(1) gauge theory. The unique aspect
here is that because of the
non-bipartite lattice all the $b_i$ carry the same sign of gauge
charge (as opposed to the staggered signs on bipartite lattices), and
 thus because of the structure of the saddle point there is a charge-2
 Higgs field coupled to the U(1) gauge theory. As originally discussed
 in seminal work such a Higgs phase leaves behind a
$\mathbb{Z}_2$ gauge theory and a topological phase~\cite{fradkin1979:higgs}. This line of
argument was used previously to establish emergent $\mathbb{Z}_2$ gauge
structures in large-$N$ expansions~\cite{read1991:spN}. We thus conclude that for $\kappa<\kappa_c$ the spin model
will be in a $\mathbb{Z}_2$ quantum spin liquid phase. This suggests
by continuity that the liquid like phase observed in our original
model, Eq.~(\ref{eqn:proj}) [the $n_b=1$ limit of Eq.~(\ref{eq:Hb})]
is also in this interesting phase. We test this conjecture below.

It is also possible to take a direct large-$N$ limit of Eq.~(\ref{eqn:proj})
(i.e. holding $n_b=1$ fixed). Analogous to work on SU($N$) models on
bipartite lattices~\cite{read1989:nucphysB} we obtain as an effective theory, a quantum dimer model on the Kagome lattice,
where the spin wavefunction is obtained by replacing each dimer with
$|S\rangle$ of Eq.~(\ref{eqn:sonsing}). At $N=\infty$ all dimer
coverings are degenerate and $1/N$ corrections introduce dynamics into
the quantum dimer model. In this limit it is clear that quadrupolar
order is absent, consistent with our numerical findings.  While the quantum dimer model so obtained is not
generally solvable, it is plausible that for
$N\gg 1$, our model  Eq.~(\ref{eqn:proj}) ends up
in the same ${\mathbb Z}_2$ spin liquid phase as an exactly solved
Kagome quantum dimer
model~\cite{misguich2002:kqdm}, since the topological spin liquid is
expected to be stable to all small deformations of the
Hamiltonian. This limit suggests that at $n_b=1$ the model remains in
a liquid state for arbitrary large $N$, as shown in Fig.~\ref{fig:nbNpd}.

\begin{figure}[t]
\centerline{\includegraphics[angle=0,width=1.0\columnwidth]{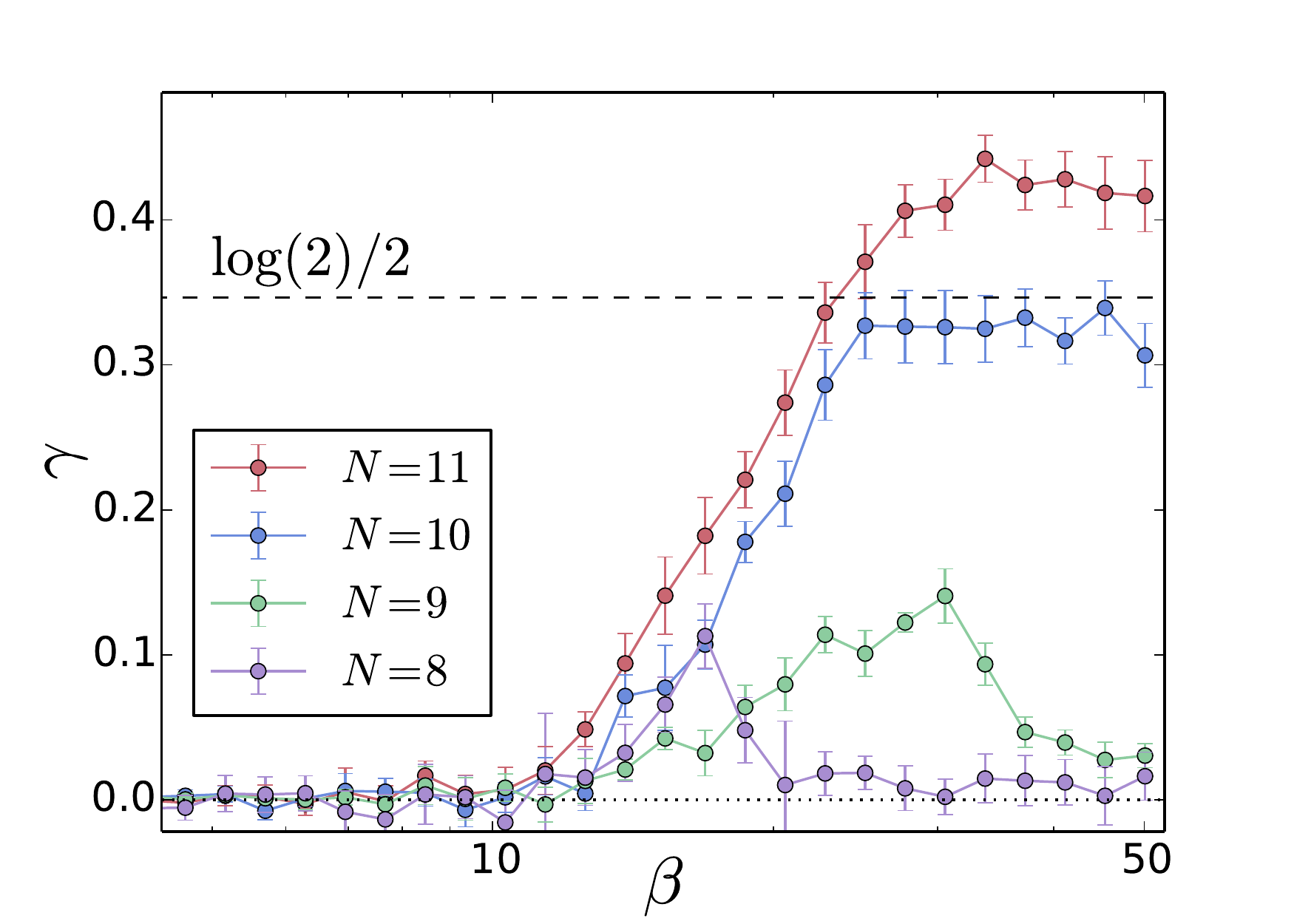}}
\caption{The topological entanglement entropy $\gamma$  of the model
  Eq.~(\ref{eqn:proj}) for various values of
  $N$ as a function of $\beta$ for system size $L=8$.  A $T=0$ quantized
  value of $\gamma$ is indicative of topological order which clearly
  manifests itself for $N\geq 10$. For $N\leq 9$, the vanishing of $\gamma$
is consistent with the appearance of quadrupolar order, see Fig.~\ref{fig:RSvsN}.}
\label{fig:L8}
\end{figure}

{\em Entanglement:} Having presented circumstantial evidence for a
spin liquid in our model Eq.~(\ref{eqn:proj}) from
large-$N$ expansions, we return to numerical simulations to
provide direct evidence for the $\mathbb{Z}_2$ quantum spin
liquid phase.  We carry out
measurements of the topological entanglement entropy (TEE), which has
been a fruitful tool to detect topological order numerically
\cite{furukawa2007:tee,isakov2011:TEEsl,grover2013:tee,jiang2012:tee}.  In phases
with topological order the TEE appears as a universal negative 
contribution to the entanglement entropy
\cite{levin2006:tee,kitaev2006:tee}. With $L$ the linear size of a smooth
simply connected subsystem, for large $L$ in a thermodynamic system:
$S_L = a L-\gamma +...$, where the first term is the
so-called ``area law'' contribution with $a$ non-universal and the
second term is the universal TEE piece.  For the $\mathbb{Z}_2$ state
found in our large-$N$ study and in the Kagome quantum dimer model, it is predicted that
$\gamma=\log(2)$ in the ground state.  This expectation has been
extended to finite temperatures as well, where due to two different excitation gaps associated with $e$ and $m$ particles, $\gamma$ is predicted to show two plateaus at $\log(2)/2$ and $\log(2)$ as a function of inverse temperature \cite{Castelnovo2007:TEEfiniteT}.

In order to isolate $\gamma$ we compute the difference in entanglement
entropy of differently shaped regions \cite{levin2006:tee}, written as
$2\gamma=S^{(2)}_{\mybendr} -S^{(2)}_{\mybox} + S^{(2)}_{\mybendl} -
S^{(2)}_{\mybars}$, where $S^{(2)}_A= -\log(\mathrm{Tr}\rho^2_A)$ is
the second R\'enyi entanglement entropy and the subscript $A$ denotes
the specific subsystem.  Writing the R\'enyi entanglement entropy in
terms of replica partition functions $S^{(2)}_A=-\log(Z^{(2)}_A/Z^2)$
\cite{Calabrese2004:EEandQFT}, the Levin-Wen measurement can be
expressed as $2\gamma= \log(Z^{(2)}_{\mybox} / Z^{(2)}_{\mybendr}) -
\log(Z^{(2)}_{\mybendl} / Z^{(2)}_{\mybars})$. To compute these
partition function ratios numerically we have adapted a recently introduced
algorithm~\cite{Demidio2019:EEfromNEW} to the current problem. The method
introduces a one parameter family of partition functions $\mathcal{Z}_{AB}^{(2)}(\lambda)$ that interpolates between the two
partition functions ($Z^{(2)}_A$ and $Z^{(2)}_B$) appearing in the
ratio. In this extended ensemble the log ratio takes the form of a
$\lambda$ integral of a simple Monte Carlo estimator~\cite{suppmat:z2}. We have also tried other techniques to calculate the EE including the
energy integration method \cite{Melko2010:MIQMC} used in
\cite{isakov2011:TEEsl}, however we find the current method to be
better suited to our problem.

\begin{figure}[!t]
\centerline{\includegraphics[angle=0,width=1.0\columnwidth]{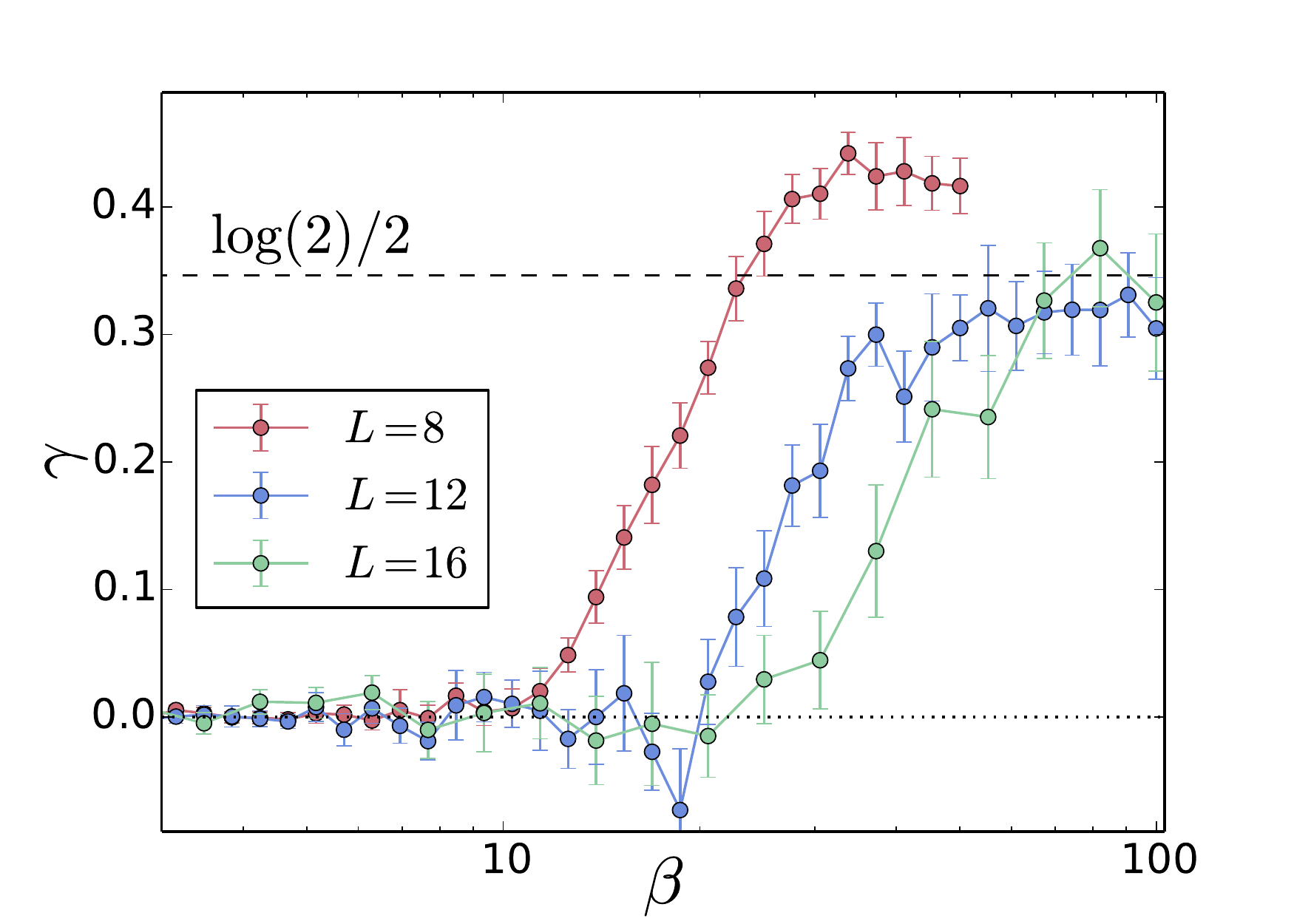}}
\caption{The topological entanglement entropy $\gamma$ as a function
  of $\beta$ for $N=11$ for $L=8,12,16$.  In the
  thermodynamic limit, we clearly see convergence to the first plateau
  of a $\mathbb{Z}_2$ spin liquid at $\log(2)/2$. Another plateau at
  $\log(2)$ is expected at still higher $\beta$.}
\label{fig:SO11}
\end{figure}

At large values of $N$ and in the quantum spin liquid phase at the low
temperatures of interest, it is
difficult to efficiently sample our phase space using only traditional
QMC loop updates. To improve the quality of our entanglement data we have
incorporated annealing and replica exchange
methods~\cite{suppmat:z2}. With these improvements we are able to
measure $\gamma$ reliably at moderately low temperatures, after which we
encounter difficulties with equilibration and ergodicity.  As we shall see, this allows us to observe the first plateau
at $\log(2)/2$ but not the second plateau at $\log(2)$. 
Fig. \ref{fig:L8} shows the TEE as a function of inverse temperature $\beta$
for the SO($N$) model with $N=8$ to $N=11$ on an $L=8$ lattice.  As $T$ is
lowered, we clearly see a
pronounced signal in the TEE for $N\geq10$ near a plateau at
$\log(2)/2$. For $N\leq 9$ on the other hand $\gamma$ goes to zero in
the low temperature regime, consistent with the study of the
quadruplar order parameter show in Fig.~\ref{fig:RSvsN}. Interestingly, even though for $L=8$ the difference
region in each ratio contains only $3 \times 2 \times 2$ we see
reasonable quantization at the first plateau.
To test how $\gamma$ scales as the system size $L$ (and
the subsystem size) is scaled up we present the
SO(11) TEE data for $L=8,12,16$ in Fig.~\ref{fig:SO11}. The data shows
clear convergence to $\log(2)/2$ as $L$ is increased.  We have also performed
measurements at lower temperatures in an effort to see the second
quantized plateau at $\log(2)$ and despite signals that are
consistent with this picture, proper equilibration here remains
challenging and will be saved for future studies.

In conclusion, we have unambiguously identified in sign-free Monte Carlo simulations, a ${\mathbb Z}_2$ quantum
spin liquid in a simple model of magnetism on a Kagome lattice. Our work
paves the way to study various interesting questions, including the
theory of phase transitions out of the QSL, the role of isolated
impurities, as well as the effect of large scale disorder in QSLs.

The computational work presented here was carried out using the XSEDE
awards TG-DMR130040 and TG-DMR140061. Financial support was received through NSF DMR-1611161.

\bibliography{z2}

\clearpage

\section{Supplemental material}

\subsection{Details on entanglement entropy measurements}

As described in the main text, we make use of the Levin-Wen construction \cite{levin2006:tee} to extract the topological entanglement entropy, which can be written as

\begin{equation}
\label{eq:TEES}
2\gamma=S^{(2)}_{\mybendr} -S^{(2)}_{\mybox} + S^{(2)}_{\mybendl} - S^{(2)}_{\mybars}.
\end{equation}
Here $S^{(2)}_A= -\log(\mathrm{Tr}\rho^2_A)$ is the second R\'enyi entanglement entropy and the subscript $A$ describes the shape of the region on an $L$ x $L$ lattice that is traced only once.  We also assume that the region size is large compared to the lattice spacing and small compared to the total system size.  Writing the R\'enyi entanglement entropy in terms of replica partition functions $S^{(2)}_A=-\log(Z^{(2)}_A/Z^2)$ \cite{Calabrese2004:EEandQFT}, the Levin-Wen measurement can be expressed as
\begin{equation}
\label{eq:TEEZ}
2\gamma= \log(Z^{(2)}_{\mybox} / Z^{(2)}_{\mybendr}) - \log(Z^{(2)}_{\mybendl} / Z^{(2)}_{\mybars}).
\end{equation}

In order to compute ratios such as these we employ the {\em equilibrium} version of a nonequilibrium method that was very recently introduced  \cite{Demidio2019:EEfromNEW}.  We have tried many different methods for these calculations, including energy integration as was used in \cite{isakov2011:TEEsl} and described in \cite{Melko2010:MIQMC}, but we find the present method to be more efficient for our model.  

In [\onlinecite{Demidio2019:EEfromNEW}] it was shown that the $\log$ ratio of partition functions $\log(Z^{(2)}_{A} / Z^{(2)}_{B})$ can be computed by introducing a weighted sum over replica partition functions $\mathcal{Z}_{AB}^{(2)}(\lambda)$ that depends on an external field $\lambda$ such that $\mathcal{Z}_{AB}^{(2)}(0)=Z^{(2)}_{B}$ and $\mathcal{Z}_{AB}^{(2)}(1)=Z^{(2)}_{A}$.  In other words $\lambda$ couples to the trace topology of the replica partition functions.  If we assume that $B$ is a subset of $A$, we can write $\mathcal{Z}_{AB}^{(2)}(\lambda)$ explicitly as

\begin{equation}
\label{eq:Zlam}
\mathcal{Z}_{AB}^{(2)}(\lambda) = \sum_{C \subseteq A-B} \lambda^{N_C}(1-\lambda)^{N_{A}-N_{B}-N_C}Z^{(2)}_{B+C}.
\end{equation}
Here $C$ is summed over all all proper subsets of the set $A-B$, from the empty set $\o$ up to and including $A-B$ itself.  Here $N_A$, $N_B$, and $N_C$ are the number of sites in the sets $A$, $B$, and $C$ respectively.  Notice that when $\lambda=0$, only $C=\o$ survives and the sum equals $Z^{(2)}_{B}$.  Also, when $\lambda=1$ only $C=A-B$ survives and the sum equals $Z^{(2)}_{A}$, as intended.  The log ratio can then be computed as

\begin{equation}
\label{eq:Sdlam}
\log(Z^{(2)}_{A} / Z^{(2)}_{B}) =\int^{1}_{0} d\lambda\frac{\partial \log \mathcal{Z}^{(2)}_{AB}(\lambda)}{\partial \lambda}.
\end{equation}
This can be efficiently measured in QMC as the equilibrium average $\partial \log \mathcal{Z}^{(2)}_{AB}(\lambda) / \partial \lambda = \frac{\langle N_C \rangle_{\lambda}}{\lambda (1-\lambda)} - \frac{N_A - N_B}{1-\lambda}$, where the average is taken in the $\mathcal{Z}^{(2)}_{AB}(\lambda)$ ensemble at a fixed value of $\lambda$.

\begin{figure}[h]
\centerline{\includegraphics[angle=0,width=1.0\columnwidth]{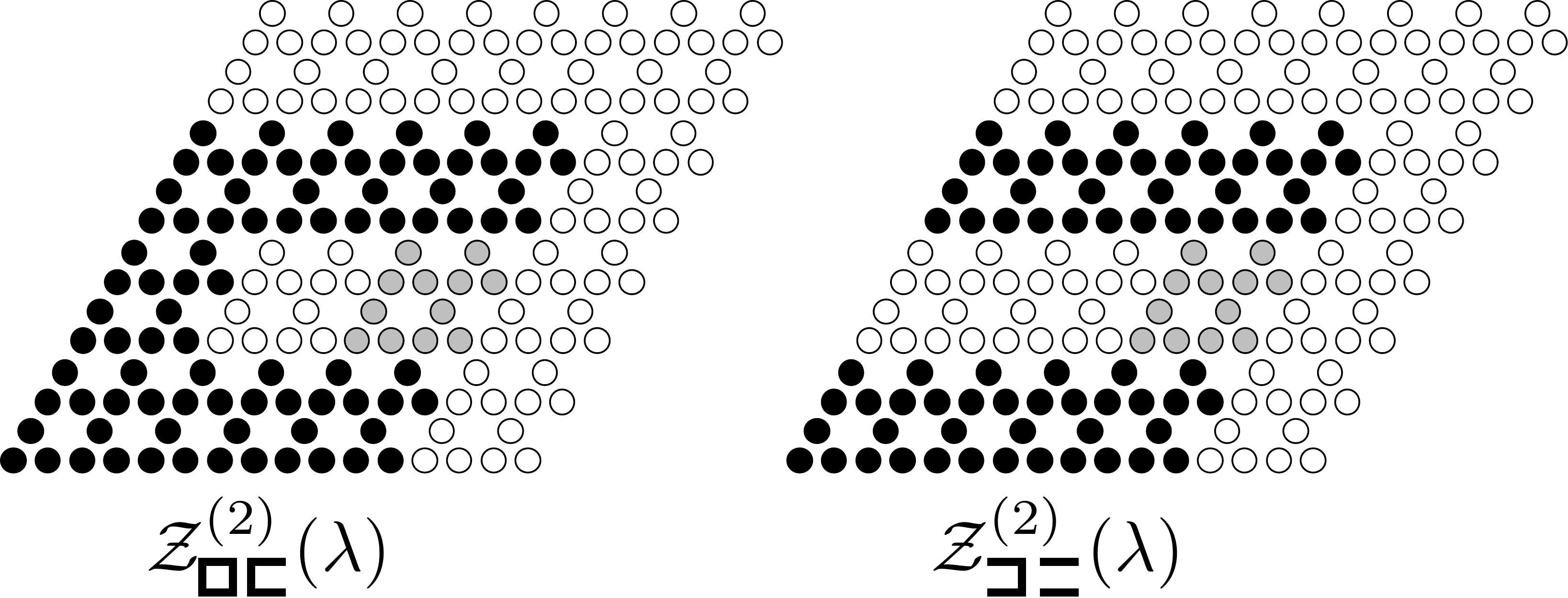}}
\caption{The two types of Levin-Wen geometries used to compute $\gamma$ in this work, here shown for an $L=8$ system.  Black sites are traced once, white sites are traced twice and grey sites independently fluctuate their trace topology according to the value of $\lambda$.  The QMC measures the average number of single trace spins in the grey region, $\langle N_C \rangle_{\lambda}$.}
\label{fig:Z1Z2}
\end{figure}
\begin{figure}[h]
\centerline{\includegraphics[angle=0,width=1.0\columnwidth]{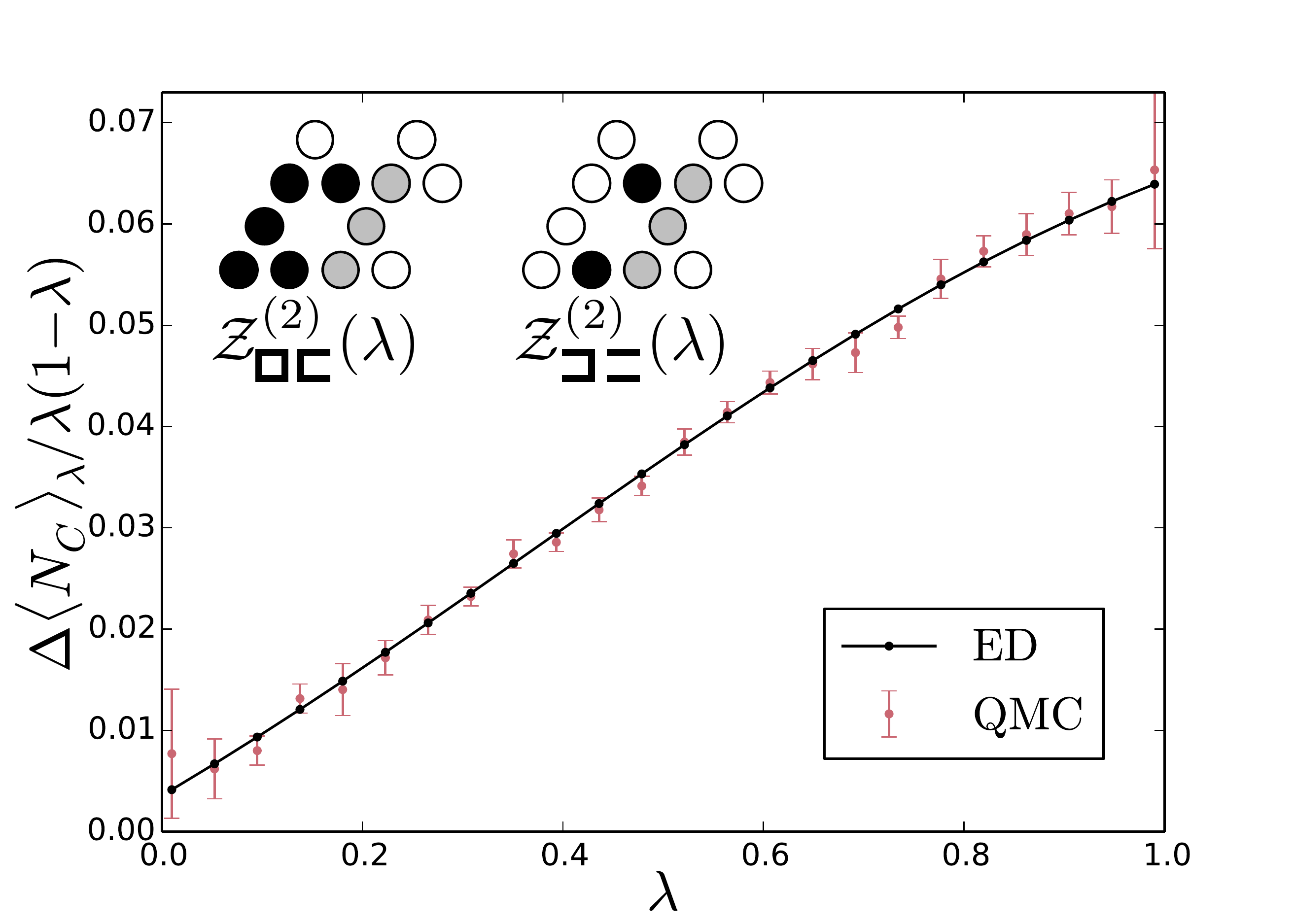}}
\caption{A comparison of the QMC measurements with values obtained by exact diagonalization.  Here we consider an $N=2$, $L=2$ system with the geometries as pictured.  $\Delta \langle N_C \rangle_{\lambda}$ is computed as the difference in the average number of single trace spins in the grey regions as a function of $\lambda$.  Exact values are obtained by diagonalizing the reduced density matrix for all bipartitions in Eq. (\ref{eq:Zlam}) with the proper $\lambda$ weight factors.  The integral under the curve gives the Levin-Wen measurement for $\gamma$, which for the $N=2$ case here would give zero in the thermodynamic limit.}
\label{fig:QMCvsED}
\end{figure}
In Fig. \ref{fig:Z1Z2} we show the two different types of $\mathcal{Z}^{(2)}_{AB}(\lambda)$ geometries that are used to compute Eq. (\ref{eq:TEEZ}) for an $L=8$ system.  In these pictures black sites are traced once, white sites are traced twice, and grey sites can fluctuate independently between a single and double trace according to the value of $\lambda$.  When $\lambda=0(1)$ all of the spins in the grey region are traced twice (once), respectively.  The QMC simply measures $\langle N_C \rangle_{\lambda}$, or the average number of spins in the grey region that are traced once for a fixed value of $\lambda$ in equilibrium.

In this framework the Levin-Wen measurement takes on a simple and intuitive form, we can write Eq. \ref{eq:TEEZ} as
\begin{equation}
\label{eq:LWlam}
2\gamma =\int^{1}_{0} d\lambda\frac{\Delta \langle N_C \rangle_{\lambda}}{\lambda (1-\lambda)},
\end{equation}
where $\Delta  \langle N_C \rangle_{\lambda}$ is the difference in the average number of single trace spins in the grey region for the two different Levin-Wen geometries in Fig. \ref{fig:Z1Z2}.

In order to measure $\gamma$ as a function of temperature, we perform thermal annealing from randomly initialized configurations at high temperature and slowly cool down to low temperature.  This is done for 24 different values of $\lambda$ for both $\mathcal{Z}^{(2)}_{\mybox \, \mybendr} (\lambda)$ and $\mathcal{Z}^{(2)}_{\mybendl \, \mybars} (\lambda)$.  On regular intervals the thermal annealing schedule is paused and measurements of $\langle N_C \rangle_{\lambda}$ are made for each ratio.  To improve measurement statistics we employ replica exchange (within each Levin-Wen geometry separately) as a function of $\lambda$, where neighboring configurations can swap $\lambda$ values probabilistically.  Since traditional QMC loop updates are extremely inefficient deep in the spin liquid phase, we compute our QMC averages over independent thermal annealing realizations.

\subsection{QMC versus exact diagonalization}

Here we compare our QMC method against exact results on a 2 x 2 lattice for $N=2$.  We have tried to make the comparison as close as possible to the measurement required to compute the topological entanglement entropy.  We have therefore chosen the Levin-Wen geometries as depicted in Fig. \ref{fig:QMCvsED}.  These are used to compute $\Delta \langle N_C \rangle_{\lambda} / \lambda (1-\lambda)$ shown in the main plot, where we see perfect agreement between the QMC and ED.

\subsection{TEE Equilibration}

\begin{figure}[h]
\centerline{\includegraphics[angle=0,width=1.0\columnwidth]{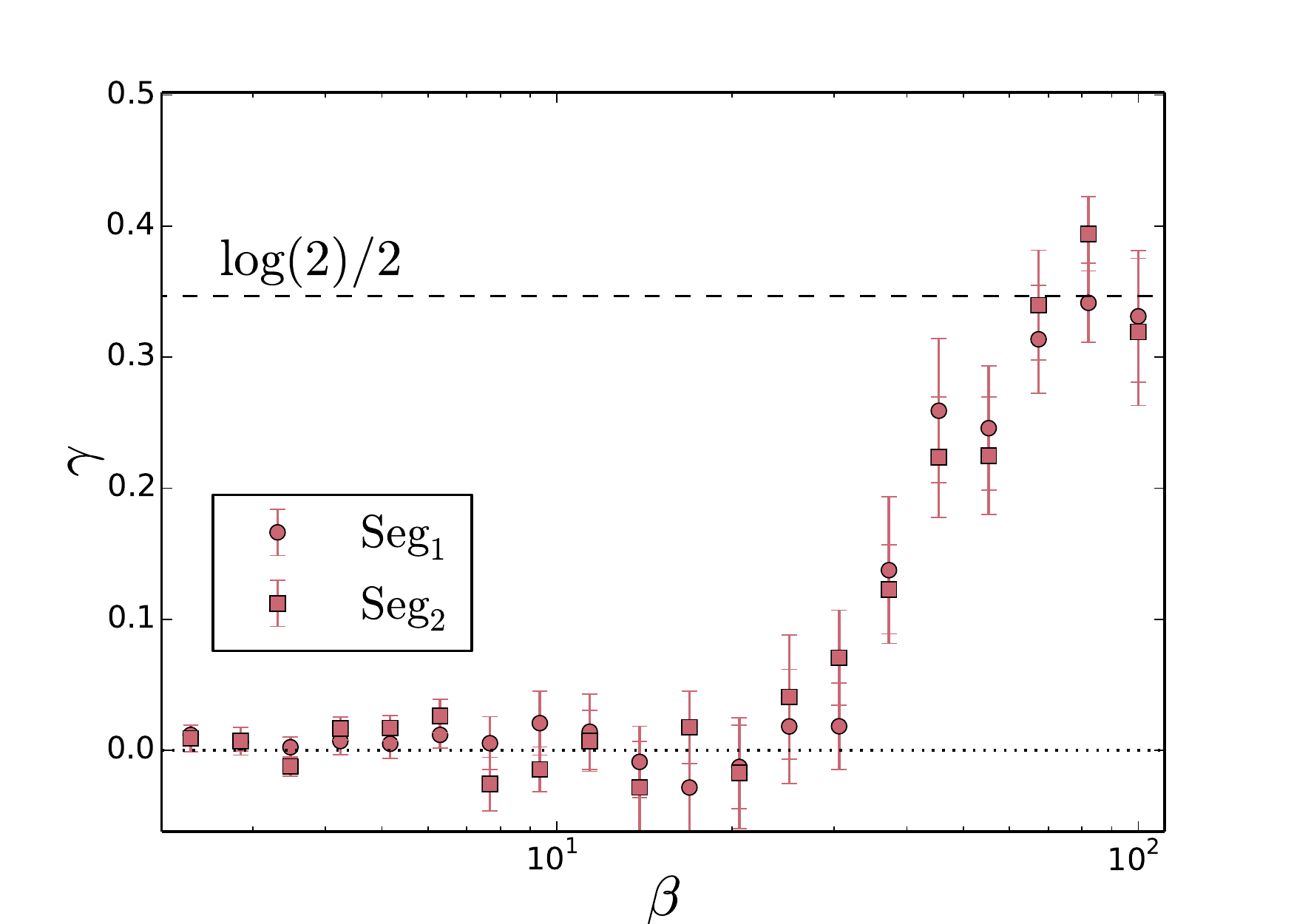}}
\caption{The first and second measurement segments for the $L=16$, SO(11) system presented in the main text.  Agreement between the segments is an indication of sufficient equilibration.}
\label{fig:L16EQL}
\end{figure}

Our thermal annealing schedule consists of 2000 equilibration sweeps at each value of $\beta$ on a fine grid of points that are equally spaced on a $\log$ scale (a geometric progression with 1400 points between $\beta=0.1$ and $\beta=100$).  The data presented in this work is generated by pausing the thermal annealing schedule at regular intervals to make measurements.  At each measured value of $\beta$ we perform 1000 equilibration sweeps, then 20000 measurement sweeps followed by another 1000 equilibration sweeps and another 20000 measurement sweeps.  We can separately average measurements from the first and second segments, where we expect statistical agreement in the case of properly equilibrated configurations.  This is shown in Fig. \ref{fig:L16EQL}, where agreement is found for the $L=16$ SO(11) system.  The data presented in the main text is the average over both measurement segments.

\subsection{TEE density}

\begin{figure}[h]
\centerline{\includegraphics[angle=0,width=1.0\columnwidth]{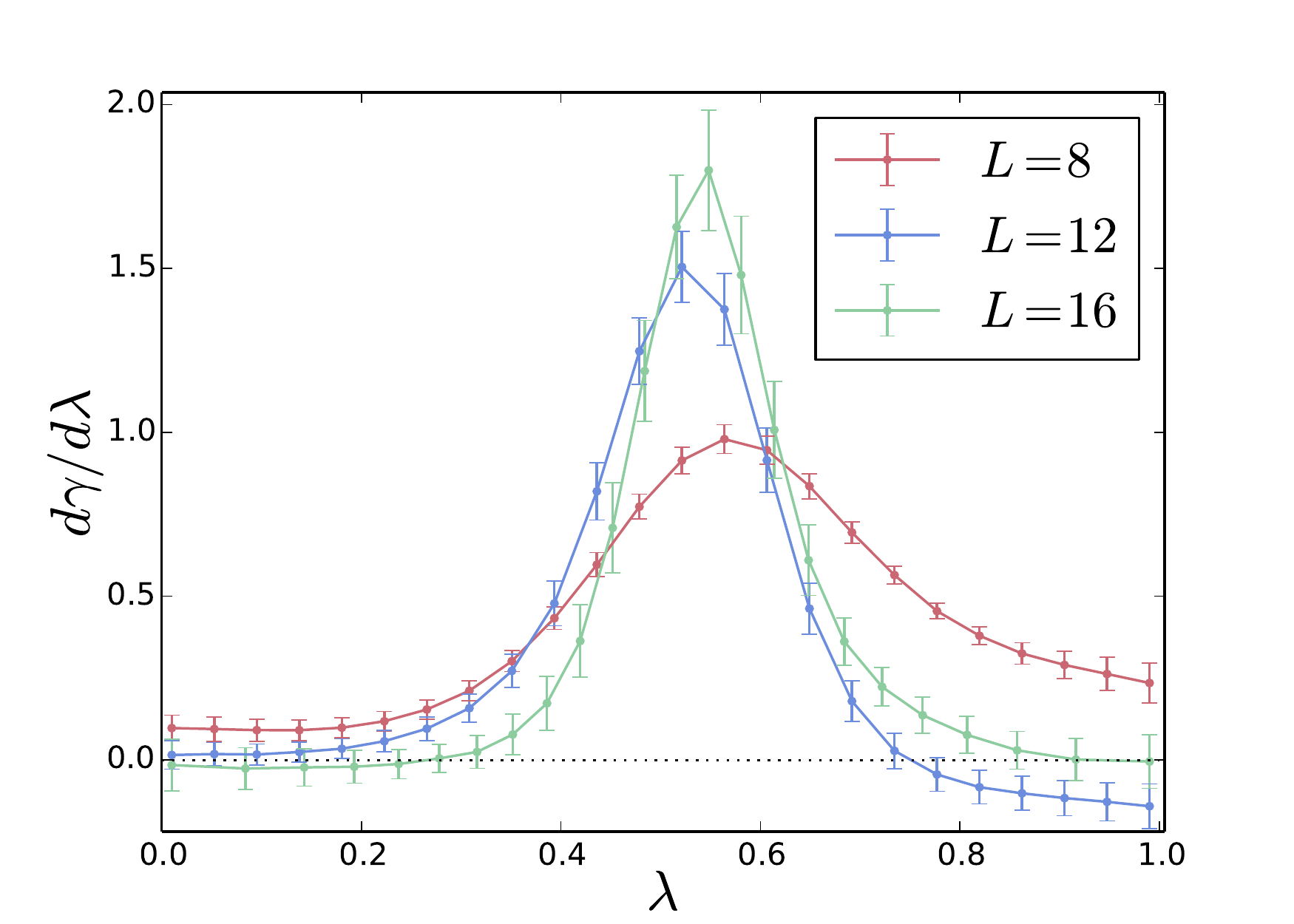}}
\caption{The topological entanglement entropy density $d\gamma/d\lambda = \Delta \langle N_C \rangle_{\lambda} / 2\lambda (1-\lambda)$ as a function of $\lambda$ at the largest values of $\beta$ used in the main text for SO(11).  The area under the curves gives $\gamma$, which receives most of the contribution near $\lambda=0.5$ and negligible contribution near the extremes for sufficiently large system sizes.}
\label{fig:TEEden}
\end{figure}

Finally it is interesting to look at the behavior of  $d\gamma/d\lambda = \Delta \langle N_C \rangle_{\lambda} / 2\lambda (1-\lambda)$ that,  when integrated from $\lambda=0$ to $\lambda=1$, gives the Levin-Wen topological entanglement entropy values presented in the main text.  This is shown in Fig. \ref{fig:TEEden} for SO(11) $L=8,12,16$ at the largest values of $\beta$ presented in the main text.  We can see that the largest contribution to $\gamma$ comes near $\lambda=0.5$ and tapers off to zero near the extremes for sufficiently large system sizes where the quantized value $\log(2)/2$ is observed.

\end{document}